\documentclass{epl}

\newcommand{\gap}[1]{2\Delta_{#1}/\ab{k}_\ab{B} T_{\ab{c}}}
\newcommand{\CsurT}{C / t \gamma_{\ab{n}} T_{\ab{c}}}
\newcommand{\kB}{\ab{k}_\ab{B}}

\title{Phenomenological two--gap model for the specific heat of  MgB$_2$}
 \shorttitle{Phenomenological two--gap model for \chem{MgB_2}}
\author{F.~Bouquet\inst{1,2} \and  Y.~Wang\inst{1} \and R.~A.~Fisher\inst{2}
 \and D.~G.~Hinks\inst{3} \and J.~D.~Jorgensen\inst{3} \and A.~Junod\inst{1}, \and
 N.~E.~Phillips\inst{2}}
 \shortauthor{F. Bouquet \etal}
 \institute{
  \inst{1} D\'epartement de Physique de la Mati\`ere Condens\'ee, Universit\'e de
Gen\`eve -- CH--1211 Gen\`eve 4 (Switzerland)\\
  \inst{2} Lawrence Berkeley
National Laboratory and Department of Chemistry, University of
California -- Berkeley, CA 94720 (USA)\\
  \inst{3}  Materials Science Division,
Argonne National Laboratory -- Argonne, IL 60439 (USA)}


\begin{document}

\maketitle

\begin{abstract}
We show that the specific heat of the superconductor \chem{MgB_2}
in zero field, for which significant non--BCS features have been
reported, can be fitted, essentially within experimental error,
over the entire range of temperature  to $T_{\ab{c}}$ by a
phenomenological two--gap model.  The resulting gap parameters
agree with previous determinations from band--structure
calculations, and from various spectroscopic experiments. The
determination from specific heat, a bulk property, shows that the
presence of two superconducting gaps in \chem{MgB_2} is a volume
effect.
\end{abstract}

The  discovery of superconductivity in \chem{MgB_2}
\cite{Nagamatsu} raised the questions of its nature and the origin
of its relatively high transition temperature $T_{\ab{c}} \sim
40$~K. Specific heat ($C$) is a powerful tool to aid in answering
these questions and, more generally, to provide information on the
thermodynamics of the transition. Several groups have reported
such measurements on \chem{MgB_2}
\cite{Budko,Kremer,Waelti,Wang,Bouquet,Yang,JunodNarlikar,%
FisherNarlikar,Marcenat}.
It is now established that $C$
 significantly deviates from the standard BCS behaviour.
First, a large excess in $C$ is observed in the vicinity of
$T_{\ab{c}} / 4$
\cite{Wang,Bouquet,Yang,JunodNarlikar,FisherNarlikar}. Second, an
exponential fit of $C(T)$ in the region $T \ll T_{\ab{c}}$
indicates a gap ratio $\gap{0}$ only one--quarter to one--third of
the isotropic BCS value \cite{Bouquet,Yang,FisherNarlikar}. This
excess was interpreted as a possible sign of a second
superconducting gap, whose existence is predicted by
band--structure calculations \cite{Shulga,Kortus,Liu}. The
specific heat near $T_{\ab{c}}$ is puzzling also with the jump
$\Delta C$ at $T_{\ab{c}}$  consistently smaller than the BCS
weak--coupling lower bound. In this Letter, we present an
empirical two--gap model that fits the experimental data over the
whole range of temperature to $T_{\ab{c}}$. This model resolves
the apparent contradiction between different analyses of the
specific heat, and relevant parameters show good agreement with
determinations based on independent experiments.

We focus on two sets of specific--heat data obtained independently
in two different laboratories. Experimental methods and
results have been described elsewhere \cite{Wang,Bouquet,%
JunodNarlikar,FisherNarlikar}. The unusual excess specific heat at
$ \sim T_{\ab{c}} / 4$,  which denotes the presence of excitations
within the main gap, is a consistent feature that is common to
different samples and different techniques. These measurements
also give similar values for the normal--state contribution, with
a coefficient of the linear term $\gamma_{\ab{n}} \sim
2.65(15)$~mJ mol$^{-1}$ K$^{-2}$, and satisfy the criterion of the
normal-- and superconducting--state entropy being equal at
$T_{\ab{c}}$. However, detailed results, such as the height and
the width of the jump $\Delta C$ at $T_{\ab{c}}$, are
sample--dependent. The sample of Ref.~\cite{Bouquet} was a powder
of isotopically pure \chem{Mg^{11}B_2} embedded in GE7031 varnish,
whereas the sample of Ref.~\cite{Wang} was a sintered commercial
powder. A third sample prepared from Mg and B by high--pressure
techniques gave similar results \cite{JunodNarlikar}. The
electronic part of the specific heat was determined by subtraction
of the normal--state data, obtained either at fields of 14 or 16~T
in Ref.~\cite{Wang}, or with a short extrapolation of the 9~T data
in Ref.~\cite{Bouquet}. We refer to the original articles for
details.

Although the low--$T$ behaviour of the specific heat data in the
earlier studies \cite{Wang,Bouquet} definitely pointed to the
presence of excitations with a characteristic energy smaller than
the BCS gap $\Delta_{BCS} = 3.53 \kB T_{\ab{c}}$, it was not clear
whether this was due to a continuous, but extreme, distribution of
the gap resulting from anisotropy, or two discrete values of the
gap closing at the same temperature $T_{\ab{c}}$, with possible
anomalous temperature dependence at some intermediate temperature.
Furthermore, it was not clear whether these models could account
for the specific heat over the whole range of temperature to
$T_{\ab{c}}$. We present here a simple empirical model, based on
the existence of two discrete gaps $\Delta_1$ and $\Delta_2$ at $T
= 0$, both closing at $T_{\ab{c}}$. In order to calculate their
respective contributions, we first consider the case of a single
gap $\Delta_0$, following the method developed by Padamsee {\em et
al.}, and generally referred to as the $\alpha$--model
\cite{Padamsee}. The ratio $\gap{0}$ (3.53 in the BCS theory) is
not fixed, but is considered to be a fitting variable. The
temperature dependence is taken to be the same as in the BCS
theory, {\em i.e.}{} $\Delta(t) = \Delta_0 \delta(t)$, where
$\delta(t)$ is the normalised BCS gap at the reduced temperature
$t = T/T_{\ab{c}}$ as tabulated by M\"uhlschlegel
\cite{Muelschlegel}. The thermodynamic properties, entropy ($S$)
and $C$, can be calculated as appropriate for a system of
independent fermion quasiparticles:
  \begin{equation}
  \label{EQentropy}
  \frac{S}{\gamma_{\ab{n}} T_{\ab{c}}} = -\frac{6}{\pi^2} \frac{\Delta_0}{\kB T_{\ab{c}}}
  \int_0^\infty \left[ f \ln f + (1 - f) \ln (1 - f)\right]\,\upd y\,
  , \qquad \frac{C}{\gamma_{\ab{n}} T_{\ab{c}}} = t \frac{\upd(S / \gamma_{\ab{n}} T_{\ab{c}})}{\upd t}\,
  ,
  \end{equation}
where $f = [\exp(\beta E) + 1]^{-1}$ and $\beta = (\kB T)^{-1}$.
The energy of the quasiparticles is given by $E = [\varepsilon^2 +
\Delta^2(t)]^{0.5}$, where $\varepsilon$ is the energy of the
normal electrons relative to the Fermi surface. The integration
variable is $y = \varepsilon / \Delta_0$.

\begin{figure}
 \onefigure{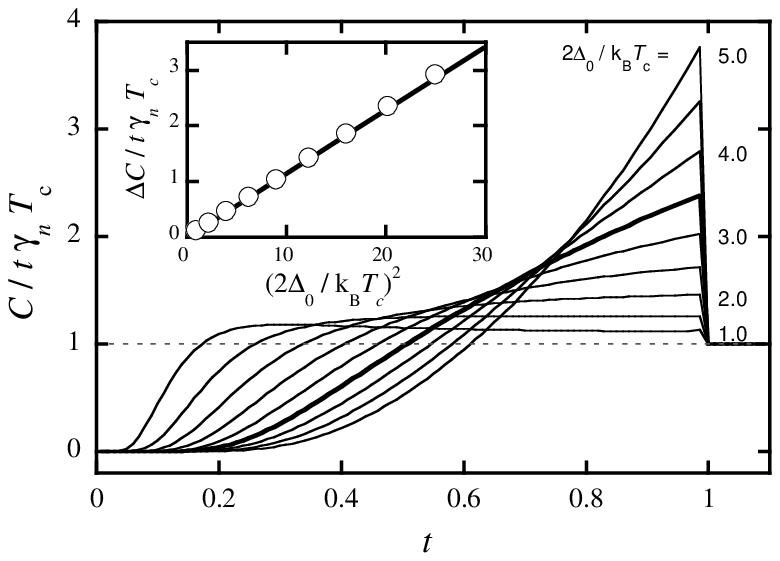}
 \caption{$\CsurT$ {\em vs.} $t$ according to the $\alpha$--model for $\gap{0} =$ 1.0,
1.5, 2.0, 2.5, 3.0, 3.5  (thick line, BCS), 4.0, 4.5, 5.0. Inset:
specific--heat jump at $T_{\ab{c}}$ {\em vs.} $(\gap{0})^2$.}
 \label{FIGalpha}
\end{figure}

The fit of experimental data for \chem{MgB_2} leads to very low
values of $\gap{0}$ for one of the gaps, substantially less than
3.53 (see below). The $\alpha$--model was devised for simulation
of strong--coupling effects \cite{Padamsee}, and has usually been
applied to strong--coupling superconductors, leading to values $>
3.53$. In that case, the temperature at which the gap closes {\em
is lowered} relative to the normal BCS closing temperature by
retardation effects. Since the BCS ratio, $\gap{0} = 3.53$, is the
weak--coupling lower limit, smaller values can have no physical
meaning as measures of the strength of the coupling. (However,
anisotropy, both as theoretically studied \cite{Clem} and
experimentally observed \cite{Okamoto}, does lead to values $<
3.5$.) In the present case, as applied to a two--gap
superconductor, a small value of $\gap{0}$ has no bearing on the
strength of the coupling , but means only that the temperature at
which the small gap closes {\em is raised} relative to the normal
BCS closing temperature by coupling to a larger gap.

Figure~\ref{FIGalpha} shows the calculated $\CsurT$ for  $1 \leq
\gap{0} \leq 5$. We checked the numerical results by comparing the
data for $\gap{0} = 3.53$ with M\"uhlschlegel's tables
\cite{Muelschlegel}, and by verifying that the entropy at
$T_{\ab{c}}$ is equal to that of the normal state. The curves for
$\gap{0} \geq 3.5$ are similar to those reported in
Ref.~\cite{Padamsee}. The unusual shape of the curves for low
values of $\gap{0}$ may be understood by considering two
characteristic temperatures, $T_\Delta =  \Delta_0/(1.76 \kB )$,
and $T_{\ab{c}}$, which are equal in the BCS limit, but  which are
independent in the present model:
\begin{itemize}
 \item For $T \ll T_\Delta$, the thermal energy is too small for many quasiparticles
to be excited across the gap. Only the tail of the statistical
distribution contributes, so that the electronic specific heat
follows an exponential behaviour approximately, similar to that of
a semiconductor.
 \item Above $T \approx T_\Delta
< T_{\ab{c}}$, the temperature is high enough to excite most of
the quasiparticles across the gap. The specific heat approaches
that of the normal state, although the system is still
superconducting.
 \item At $T = T_{\ab{c}}$, the gap closes. If $T_{\ab{c}} \gg T_\Delta$, {\em
i.e.}{} if the gap is small compared to the thermal energy at
$T_{\ab{c}}$, only a small change occurs in the number of excited
quasiparticles. The BCS ground state is essentially empty. As a
consequence, the specific--heat jump is small.
\end{itemize}

\begin{figure}
 \onefigure{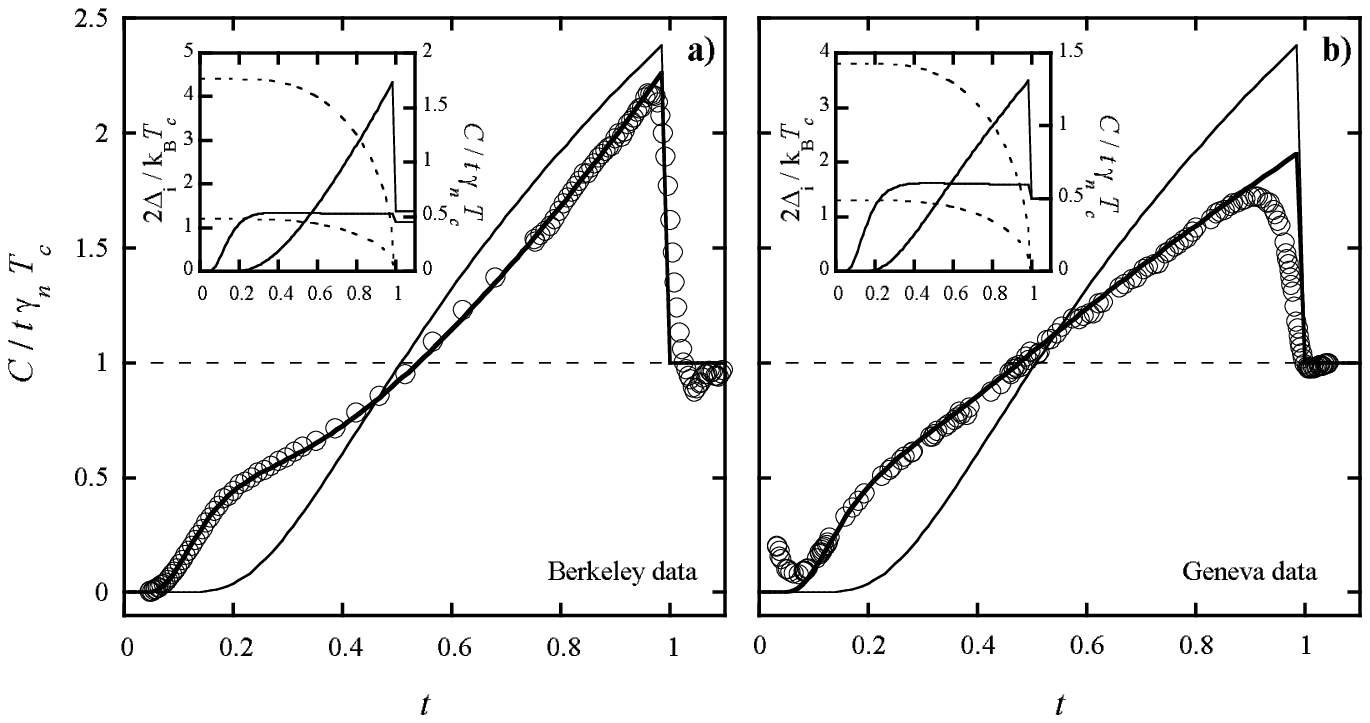}
 \caption{BCS normalized specific heat (thin line), experimental
data ($\circ$), and two--gap fits (thick lines), versus the
reduced temperature $t$. (a) data from
Ref.~\protect\cite{Bouquet}; (b) data from
Ref.~\protect\cite{Wang}. Insets: gaps $\gap{1}$ and $\gap{2}$
versus $t$ (dotted lines), and partial specific heat of both bands
(full lines). Parameters obtained from the fits are given in
Table~\ref{TAB}.}
 \label{FIGfit}
\end{figure}

The smaller the gap, the closer the $\CsurT$ curve approaches the
normal--state line, and the smaller the $\Delta C$ at the
transition. We verify numerically the relation between the gap and
the jump, $\Delta C = \kB N(0) / (\kB T_{\ab{c}})^2(\upd
\Delta^2/\upd \beta) \propto \Delta_0^2$ (inset of
Fig.~\ref{FIGalpha}) \cite{BCS}. This quadratic dependence holds
only because the variation of the normalised gap with $t$ is
common to all curves.

In a two--band, two--gap model, the total specific heat can be
considered as the sum of the contributions of each band calculated
independently according to eq.~(\ref{EQentropy}) if interband
transitions due to scattering by impurities or phonons can be
neglected. Each band is characterised by a partial Sommerfeld
constant $\gamma_i$, with $\gamma_1 + \gamma_2 = \gamma_{\ab{n}}$.
$C$ data are fitted with three free parameters, the gap widths
$\Delta_1$ and $\Delta_2$, and the relative weights $\gamma_1 /
\gamma_{\ab{n}} \equiv x$ and $\gamma_2 / \gamma_{\ab{n}} \equiv 1
- x$. Figure~\ref{FIGfit} shows the data (circles) and  the fit
(thick line), compared to the BCS specific heat (thin line).
Insets show the gap functions, and the various contributions to
the total electronic specific heat. The latter curves show
evidence of weak correlation between the fitting parameters; the
low--temperature excess is related to $\Delta_2$, whereas the jump
at $T_{\ab{c}}$ is due essentially to the $\Delta_1$ component.
Numerical results are given in Table~\ref{TAB}.

\begin{figure}
 \twofigures{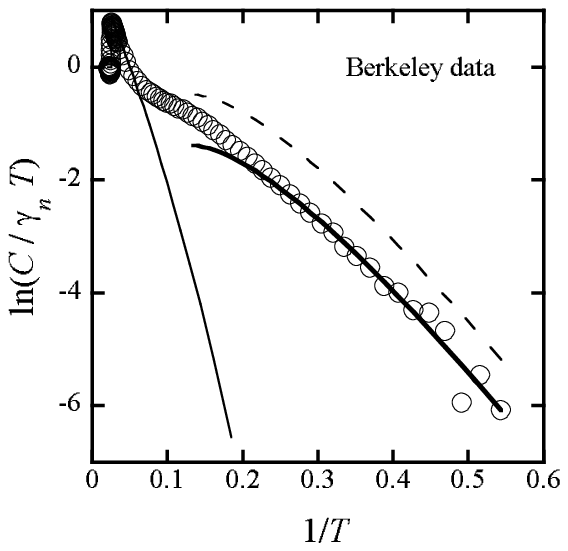}{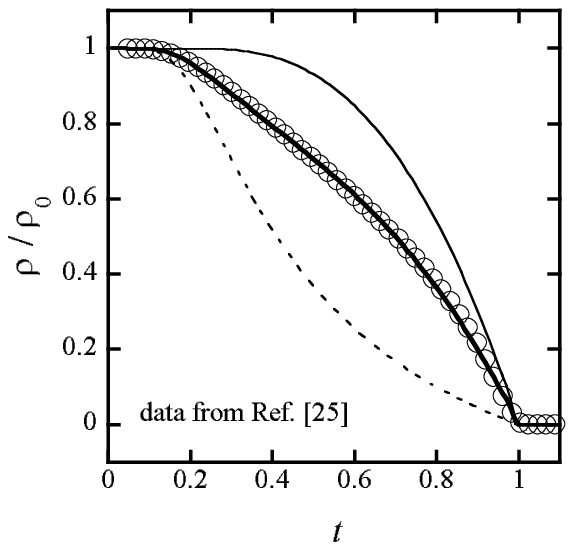}
 \caption{Semi--logarithmic plot of the electronic specific heat versus $1/T$.
Dashed line: asymptotic curve, eq.~(\protect\ref{EQkresin}) with
$\gamma = \gamma_{\ab{n}}$; thick line:
eq.~(\protect\ref{EQkresin}) with $\gamma = 0.4 \gamma_{\ab{n}}$
(see text); thin line:  standard BCS curve, also shown in
Fig.~\protect\ref{FIGfit}; ($\circ$), data from
Ref.~\protect\cite{Bouquet}.}
 \label{FIGlowt}
 \caption{Superfluid fraction versus reduced temperature. Thin
line: contribution of $\Delta_1$; dotted line: contribution of
$\Delta_2$; thick line: full two--gap fit; ($\circ$): data
obtained from measurements of the penetration depth presented in
Ref.~\protect\cite{Carrington}. Fitted parameters are given in
Table~\protect\ref{TAB}.}
 \label{FIGtony}
\end{figure}

In spite of its limitations, this empirical model fits the
measured specific heat  well over the whole range of $T$  to
$T_{\ab{c}}$. The sample dependence of the results is reasonably
low, and may reflect metallurgical differences. The larger value
of $\Delta_1$ for the sample of Fig.~\ref{FIGfit}a (isotopically
pure \chem{Mg^{11}B_2} powder \cite{Bouquet}) reflects a sharper
jump and a steeper slope just below $T_{\ab{c}}$ compared to the
sample of Fig.~\ref{FIGfit}b (\chem{MgB_2} sinter \cite{Wang}). On
average, $\gap{1} \sim 4.0$ and $\gap{2} \sim 1.2$, with
approximately equal weights.

Moreover, the fitted parameters are qualitatively and
quantitatively comparable with independent determinations from
other sources. They are consistent with band--structure
calculations
\cite{Liu} and spectroscopic measurements \cite{Chen,Tsuda,%
Giubileo,Szabo,Laube}, which  report the presence of two gaps, the
smaller gap having approximately one--third the BCS value and the
larger gap being slightly greater than the BCS value
(Table~\ref{TAB}). We emphasise that $C$, a thermodynamic
property, probes the whole volume, whereas spectroscopic
measurements are more sensitive to surface conditions.

The relative weights (1:1, {\em i.e.} $x \sim 0.5$) are consistent
with the calculations of Ref.~\cite{Liu}. Liu {\em et al.}{}
attribute the larger gap $\Delta_1$ to particular 2D sheets of the
Fermi surface, whereas the smaller gap $\Delta_2$ is associated
with 3D sheets. Using partial densities of states and de Haas--van
Alphen mass renormalizations, the weight of the smaller gap is
evaluated as $x \sim 0.47$, and $1 - x \sim 0.53$ for the larger
one. The agreement with the two--gap model fits is remarkable.

\begin{table}
\caption{Gap ratios $\gap{1}$, $\gap{2}$ and weights $x$ as
determined by the two--gap model (lines 1--4) and by different
techniques (lines 5--10).} \label{TAB}
\begin{center}
\begin{largetabular}{ccccc}
 Ref. & Technique & $\gap{1}$ & $\gap{2}$ & $x:(1-x)$ \\
 \hline
 \cite{Bouquet} & specific heat & 4.4 & 1.2 & 55\% : 45\% \\
 \cite{Wang} & specific heat & 3.8 & 1.3 & 50\% : 50\% \\
 \cite{JunodNarlikar} & specific heat & 3.9 & 1.3 & 50\% : 50\% \\
 \cite{Carrington} & penetration depth & 4.6 & 1.6 & 60\% : 40\% \\
 \cite{Chen} & Raman & 3.7 & 1.6 &  \\
 \cite{Tsuda} & photoemission & 3.6 & 1.1 &  \\
 \cite{Giubileo} & tunneling & 4.5 & 1.9 &  \\
 \cite{Szabo} & point--contact spectroscopy & 4.1 & 1.7 &  \\
 \cite{Laube} & point--contact spectroscopy & 4.2 & 1.0 &  \\
 \cite{Liu} & band structure & 4.0 & 1.3 & 53\% : 47\% \\
\end{largetabular}
\end{center}
\end{table}

The present two--gap model reconciles the apparently conflicting
results of Ref.~\cite{Kremer} and \cite{Yang,Bouquet}. By fitting
their  specific heat data close to $T_{\ab{c}}$, Kremer {\em et
al.}{} \cite{Kremer}  concluded that their data was consistent
with a medium-- to strong--coupling $\gap{0} \sim 4.2$. However,
the fitted value of $\gamma_{\ab{n}}$ at $T_{\ab{c}}$ was 1.1~mJ
mol$^{-1}$K$^{-2}$, less than half of $\gamma_{\ab{n}}$ measured
in the normal state. Alternatively, Yang {\em et al.}{}
\cite{Yang} and Bouquet {\em et al.}{}\cite{Bouquet} fitted the
exponential decrease of the low--$T$ data and concluded that
$\gap{0} \sim 0.9$. However, the fitted value of $\gamma_{\ab{n}}$
at low $T$ was too small also, 0.7~mJ mol$^{-1}$K$^{-2}$ in
Ref.~\cite{Bouquet}. In the framework of the two--gap model, the
main contribution just below $T_{\ab{c}}$ is that of the larger
gap $\Delta_1$, with a break in the slope characteristic of
medium-- to strong--coupling, and an amplitude of $\Delta C$
determined by $\gamma_1 = x \gamma_{\ab{n}} \sim
\gamma_{\ab{n}}/2$ (insets of Fig.~\ref{FIGfit}), in qualitative
agreement with Kremer's analysis. The main contribution at $T \ll
T_{\ab{c}}$ is that of the smaller gap $\Delta_2$, with the
exponential decrease determined by $\Delta_2$, and the amplitude
by $\gamma_2 = (1-x) \gamma_{\ab{n}} \sim \gamma_{\ab{n}}/2$
(insets of Fig.~\ref{FIGfit}), again in qualitative agreement with
the analysis of Ref.~\cite{Bouquet}. The latter data are presented
below in a slightly different approach. Rather than the usual
empirical interpolation $C \propto \exp(-1.44T_{\ab{c}}/T)$, we
use the low--$T$ asymptotic formula \cite{Kresin}:
  \begin{equation}
  \label{EQkresin}
 \lim_{T \rightarrow 0} \frac{C}{\gamma T} = 3.15  \left(
 \frac{\Delta_0}{1.76 \kB T}\right)^{5/2}
 \exp\left(-\frac{\Delta_0}{\kB T}\right) \, .
  \end{equation}
In Fig.~\ref{FIGlowt}, we plot data in the form
$\ln(C/\gamma_{\ab{n}} T)$ versus $1/T$, together with the limit
given by eq.~(\ref{EQkresin}). With $\gap{0} = 0.9$ and $\gamma =
\gamma_{\ab{n}}$, eq.~(\ref{EQkresin}) overestimates the data,
although the slope determined by $\Delta_0$  is correct. With
$\gamma \cong 0.4 \gamma_{\ab{n}}$, the fit is good in the domain
where eq.~(\ref{EQkresin}) holds.

The same two--gap model can be applied to the superfluid density
$\rho$, which is given, for a single gap, by $
  \rho = 1 - 2\Delta_{0}/\ab{k}_\ab{B} T \int_0^\infty f(1-f)\,\upd y$.
The penetration depth $\lambda \propto \rho^{-1/2}$ is given in
Ref.~\cite{Carrington} and is plotted in Fig.~\ref{FIGtony},
together with a two--gap fit (thick line) and its components (full
and dotted lines). These data are not strictly bulk measurements,
but probe the sample to a typical depth  of $\lambda \cong
1800$~\AA \cite{Wang}. Nevertheless, $\lambda$ is large compared
to the typical sampling depth of many spectroscopic experiments,
which is on the scale of the coherence length $\xi \cong
50$~\AA \cite{Wang}.  The fitted parameters $\gap{i}$ and $x$ are
consistent with other determinations (Table~\ref{TAB}).

The empirical $\alpha$--model allows a quantitative comparison to
be made between different experiments and theory within a general
framework. The results are numerically consistent, and confirm the
coexistence of two gaps for the bulk sample. This situation holds
the promise of interesting single--crystals properties. Our
two--gap model is phenomenological since we {\em postulate} the
existence of the gaps, without specifying their origin. Any
theoretical approach leading to a similar average electronic
density of states would be compatible with the present results, so
that specific--heat measurements alone cannot settle in favour of
any particular microscopic model \cite{Liu,Shulga}.

Some limitations exist. First, the $\alpha$--model assumes a
BCS--like $T$--dependence of the gap. However, if the variation of
the smaller gap is reasonably smooth, the results should not
depend critically on its exact shape, since the main effect on the
specific heat occurs below $T_\Delta$ where $\Delta_2(T)$ is
expected to be essentially constant. Self--consistent calculations
of $\Delta(T)$ might lead to corrections, and more elaborate
simulations are currently under way \cite{Drechsler}. Second, we
calculate each contribution of the gaps independently and assume
that they are additive. Some coupling is present, but within the
present model, its sole effect amounts to bringing the natural
closing temperature of the smaller gap, {\em i.e.}{} $\approx
10$~K, up to $\approx 40$~K.

Our two--gap model describes only the zero--field specific heat.
As data in $H > 0$ suggest a different field dependence for each
gap, a theory of the mixed--state specific heat for a two--gap
superconductor would be most useful in extracting quantitative
information from $C(T,H)$. Indeed,  the field dependence of the
electronic contribution at low temperature is unusual. The
coefficient of the linear term in the mixed--state, $\gamma(H)$,
dramatically increases in small fields
\cite{Wang,Bouquet,Yang,JunodNarlikar,FisherNarlikar}, in a
quasi--logarithmic way \cite{JunodNarlikar}, and saturates for
fields much below $H_{\ab{c2}}$, in fact near $H_{\ab{c2}}/2$.
Moreover, the characteristic dip in $C/T$ for $0 < T < 10$~K
associated with the gap $\Delta_2$ in one of the bands vanishes by
$\approx 0.5$~T. Qualitatively, these results would seem to
indicate that a small field is able to quench the smaller gap, in
agreement with spectroscopic measurements \cite{Szabo}. The
saturation of $\gamma(H)$ much below $H_{\ab{c2}}$ suggests that a
major part of the electrons are either normal or in a gapless
state, possibly by virtue of inter--band scattering in the
presence of a normal sheet on the Fermi surface. Forthcoming
models  may have to embody the different dimensionality of the
gaps. Interesting developments are expected for the physics of the
vortex state for superconductors with such an unusual
$k$--dependent gap.

\acknowledgments

Stimulating discussions with I.~L.~Mazin, A.~Carrington,
V.~Kresin, E.~Schachinger, S.~V. Shulga and S.--L.~Drechsler are
gratefully acknowledged. This work was supported by the Fonds
National Suisse de la Recherche Scientifique and by the US
Department of Energy.

\end{document}